\documentclass[prl,twocolumn,superscriptaddress]{revtex4}
\usepackage{graphicx}
\usepackage{dcolumn}
\usepackage{amsmath,amsthm,amssymb}
\usepackage{subfigure}

\def\ep{\varepsilon}

\begin{document}
\title{Weakly turbulent instability of anti-de Sitter space}

\author{Piotr Bizo\'n}
\affiliation{Institute of Physics, Jagiellonian
University, Krak\'ow, Poland}
  \affiliation{Max-Planck-Institut f\"ur Gravitationsphysik,
Albert-Einstein-Institut, Golm, Germany}
\author{Andrzej Rostworowski}
\affiliation{Institute of Physics, Jagiellonian
University, Krak\'ow, Poland}
\date{\today}
\begin{abstract}
We study  the nonlinear evolution of a weakly perturbed anti-de Sitter (AdS) space
   by solving numerically the four-dimensional spherically symmetric Einstein-massless-scalar field equations with negative cosmological constant. Our  results suggest that AdS space is unstable under
    \emph{arbitrarily} small generic perturbations.
   We conjecture that this instability is triggered by a resonant mode mixing  which gives rise to diffusion of energy from low to high frequencies.

\end{abstract}

\maketitle

\noindent \emph{Introduction.}
 The past decade has witnessed growing interest in  spacetimes which asymptotically approach
  anti-de Sitter space (AdS) which is the unique maximally symmetric solution of the vacuum Einstein equations  with  negative cosmological constant.
 Despite this flurry of activity, motivated mainly by the AdS/CFT duality conjecture, the very basic question "Is AdS stable?"~\cite{stability} has been rarely raised (cf.~\cite{a0} for a notable exception), let alone answered. This is in stark contrast to two other maximally symmetric solutions of the vacuum Einstein equations -- Minkowski space (with zero cosmological constant) and de Sitter space (with positive cosmological constant) -- which are known to be stable under small perturbations \cite{ck,f1}. The key feature of asymptotically AdS spacetimes, distinguishing them from asymptotically flat or dS spacetimes, is the presence of a timelike boundary  at (spatial and null)  infinity where suitable boundary conditions need to be prescribed in order to make  the evolution well-defined \cite{f2}. For no-flux  boundary conditions one gets a Hamiltonian conservative system on an effectively  bounded domain \cite{boundary} from which energy cannot escape (as for waves propagating inside a perfect cavity), hence the
 asymptotic stability of AdS space is precluded while the question of its stability, as we will see below, touches upon the KAM theory for partial differential equations.
\vskip 0.15cm
\noindent\emph{Model.} In this Letter we report on numerical simulations which shed some new light on the problem of stability of AdS spacetime.
  As a simple model of asymptotically AdS dynamics we consider a self-gravitating spherically symmetric  real massless scalar field in $3+1$ dimensions whose evolution is governed by the Einstein-scalar system with negative cosmological constant $\Lambda<0$
\begin{align}\label{scalar}
& G_{\alpha\beta} + \Lambda g_{\alpha \beta} =
 8 \pi G \left(\partial_{\alpha} \phi \,\partial_{\beta} \phi - \frac{1}{2}  g_{\alpha\beta} (\partial \phi)^2\right), \\
&  g^{\alpha\beta} \nabla_{\alpha} \nabla_{\beta} \phi=0\,.
\end{align}
For the metric we assume the ansatz
\begin{equation}
\label{ads3+1:ansatz}
ds^2 = \frac {\ell^2}{\cos^2{\!x}}\left( -A e^{-2 \delta} dt^2 + A^{-1} dx^2 + \sin^2{\!x} \,  d\Omega^2\right)\,,
\end{equation}
where $\ell^2=-3/\Lambda$ and $d\Omega^2$ is the standard metric on the round unit two-sphere. The ranges of dimensionless coordinates are $-\infty<t<\infty$ (for the 'unwrapped' version of AdS considered in this paper) and $0\leq x<\pi/2$.
We assume that $A$, $\delta$, and $\phi$ are functions of $(t,x)$. Inserting the ansatz \eqref{ads3+1:ansatz} into the field equations (1-2) and introducing auxiliary variables
$\Phi= \phi'$ and $\Pi= A^{-1} e^{\delta} \dot \phi$ (hereafter overdots and primes denote derivatives with respect to $t$ and $x$, respectively), we obtain a coupled quasilinear elliptic-hyperbolic system, consisting of the wave equation
\begin{equation}
\label{ms_in_ads3+1:eq_wave}
\dot\Phi = \left( A e^{-\delta} \Pi \right)', \qquad \dot \Pi = \frac{1}{\tan^2{\!x}}\left(\tan^2{\!x} \,A e^{-\delta} \Phi \right)',
\end{equation}
and the constraints (using units where $4\pi G=1$)
\begin{align}
\label{ms_in_ads3+1:eq_00}
A' \!&= \!\frac{1+2\sin^2{\!x}} {\sin{x}\cos{x}} \, (1-A) - \sin{x}\cos{x} \, A \left( \Phi^2 + \Pi^2 \right),
\\
\label{ms_in_ads3+1:eqs_10_11}
\delta' \!&=\! -  \sin{x}\cos{x} \left( \Phi^2 + \Pi^2 \right).
\end{align}
Note that the length scale $\ell$ drops out from the equations. The pure AdS solution is given by $\phi=0, A=1, \delta=0$.
We want to solve the system (4-6) for smooth initial data with finite total mass. Smoothness at the origin implies that near $x=0$ we have the power series expansions
\begin{align}\label{x=0}
    \phi(t,x)&= f_0(t)+\mathcal{O}(x^2),\quad  \delta(t,x)= \mathcal{O}(x^2), \nonumber \\
    A(t,x)&=1+\mathcal{O}(x^2)\,,
\end{align}
where we used normalization $\delta(t,0)=0$ so that $t$ is the proper time at the center.
These expansions are uniquely determined by a free function $f_0(t)$.
Smoothness at spatial infinity and finiteness of the total mass $M$ imply that near $x=\pi/2$ we have (using $\rho=\pi/2-x$)
\begin{align}\label{pi2}
    \phi(t,x)&= f_{\infty}(t)\, \rho^3+\mathcal{O}\left(\rho^5\right),\quad
    \delta(t,x)=\delta_{\infty}(t)+ \mathcal{O}\left(\rho^6\right), \nonumber \\
    A(t,x)&=1- 2M \rho^3+\mathcal{O}\left(\rho^6\right),
\end{align}
where $M$  and free functions $f_{\infty}(t)$, $\delta_{\infty}(t)$ uniquely determine the power series expansions.
  It follows from \eqref{pi2} that for smooth initial data  there is no freedom in imposing the boundary data (this is directly related to the fact that the corresponding linearized operator is essentially self-adjoint \cite{iw}), hence the initial value problem  is well-defined (see \cite{hs} for the rigorous proof) without the need of specifying boundary data at infinity.
  \vskip 0.15cm
  \noindent\emph{Numerical results.} We solved the system (4-6) numerically  using a fourth-order accurate finite-difference code.  We used the method of
lines and a 4th-order Runge-Kutta scheme to integrate the wave equation
(\ref{ms_in_ads3+1:eq_wave}) in time, where at each step the metric
functions were updated by solving the hamiltonian constraint (\ref{ms_in_ads3+1:eq_00}) and
the slicing condition (\ref{ms_in_ads3+1:eqs_10_11}). Preservation of the momentum constraint
$\dot A + 2 \sin{x}\cos{x} \, A^2 e^{-\delta} \Phi\, \Pi =0$
was monitored to check the accuracy of the code.

Solutions shown in Figs.~1 and 2 were generated from Gaussian-type initial data of the form
\begin{equation}\label{idata}
    \Phi(0,x)=0\,,\quad \Pi(0,x)=\frac{2\ep}{\pi} \exp\left(-\frac{4\tan^2{\!x}}{\pi^2\sigma^2}\right)\,,
\end{equation}
 with fixed width $\sigma=1/16$ and varying amplitude $\ep$. For such data the scalar field is well localized in space and propagates in time as a narrow wave packet.
For large amplitudes the wave packet quickly collapses, which is signalled by the formation of an apparent horizon at a point $x_H$ where $A(t,x)$ drops to zero. As the amplitude is decreased, the horizon radius $x_H$ decreases as well and goes to zero for some critical amplitude $\ep_0$. This behavior is basically the same as in the asymptotically flat case, because for $x_H\ll \pi/2$ the influence of the AdS boundary is negligible. At criticality the $\Lambda$ term becomes completely irrelevant, hence the solution with amplitude $\ep_0$ asymptotes (locally, near the center) the discretely self-similar critical solution discovered by Choptuik in the corresponding model with $\Lambda=0$ \cite{matt}.
 For  amplitudes slightly below $\ep_0$ the wave packet travels to infinity, reflects off the boundary,  and collapses while approaching the center. Lowering gradually the amplitude we find the second critical value $\ep_1$ for which $x_H=0$. As $\ep$ keeps decreasing, this scenario repeats again and again, that is we obtain a decreasing sequence of critical amplitudes $\ep_n$ for which  the evolution, after making $n$ reflections from the AdS boundary,
 locally asymptotes Choptuik's solution. Specifically,  we verified that in each small right neighborhood of $\ep_n$  the horizon radius  scales according to the power law $x_H(\ep)\sim (\ep-\ep_n)^{\gamma}$ with $\gamma\simeq 0.37$.  Fig.~1 shows that $x_H(\ep)$ has the shape of the right continuous sawtooth curve with finite jumps at each $\ep_n$. Notice that $T(\ep_{n+1})-T(\ep_n)\approx \pi$, where $T(\ep)$ denotes the time of collapse. We stress that $x_H$ is the radius of the \emph{first} apparent horizon that forms on the $t=const$ hypersurface;  eventually all the matter falls into the black hole and the solution settles down to the Schwarzschild-AdS black hole with mass equal to the initial mass $M$ (cf. \cite{hs2}). It appears that $\lim_{n\rightarrow \infty} \ep_n=0$, indicating that there is no threshold for black hole formation, however we did not determine precise values of $\ep_n$ for $n>10$  because the computational cost of bisection increases rapidly with $n$ (since, in order to resolve the collapse, solutions have to be evolved for longer times on finer grids ).

  Let us mention that the analogous problem in $2+1$ dimensions was studied previously by Pretorius and Choptuik \cite{cp} who emphasized the challenges inherent in numerical simulations of AdS dynamics, however their analysis was primarily focused on the threshold for black hole formation  \emph{before} any reflection off the AdS boundary takes place (as for our data with amplitude $\ep_0$).
\begin{figure}
  \includegraphics[width=0.48\textwidth]{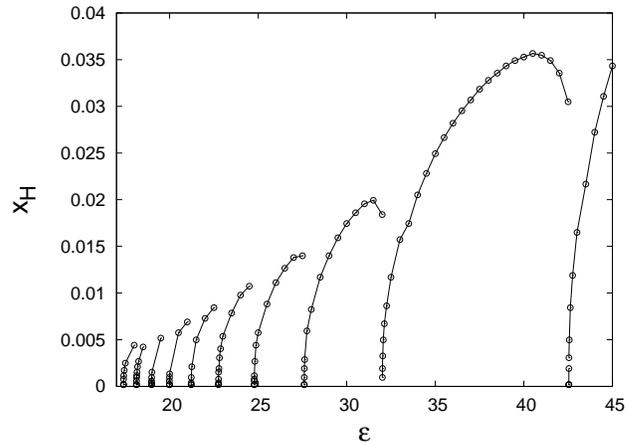}
  \caption{
  Horizon radius vs amplitude for initial data \eqref{idata}.
  The number of reflections off the AdS boundary before collapse varies from zero to nine
  (from right to left).\vspace{-0.4cm}}
  \label{fig1}
\end{figure}

 In the following we consider the development of  general (gaussian and other) small initial data,
focusing  attention  on early and intermediate pre-collapse  phases of evolution. We found that the Ricci scalar at the center, $R(t,0)=-2\Pi^2(t,0)/\ell^2-12/\ell^2$, can serve as a good indicator for the onset of instability. This quantity oscillates with  frequency $\approx 2$ (as it takes time $\approx \pi$ for the wave packet to make the round trip from and back to the center). An upper envelope of these oscillations
is shown in Fig.~2a, where several
 clearly pronounced phases of evolution can be distinguished.
During the first phase the amplitude remains approximately constant but after some time  there begins a second phase of (roughly)  exponential growth, followed by subsequent phases of steeper and steeper growth, until finally the solution collapses. We find that the time of onset of the  second phase scales as $\ep^{-2}$ (see Fig.~2b), which means that arbitrarily small perturbations eventually start growing. Note that this behavior is morally tantamount to instability of AdS space, regardless of what happens later, in particular whether the solution will collapse or not. In the remainder of this Letter we sketch a preliminary attempt to explain the mechanism of this instability in the framework of weakly nonlinear perturbation theory.
\begin{figure}
  \includegraphics[width=0.48\textwidth]{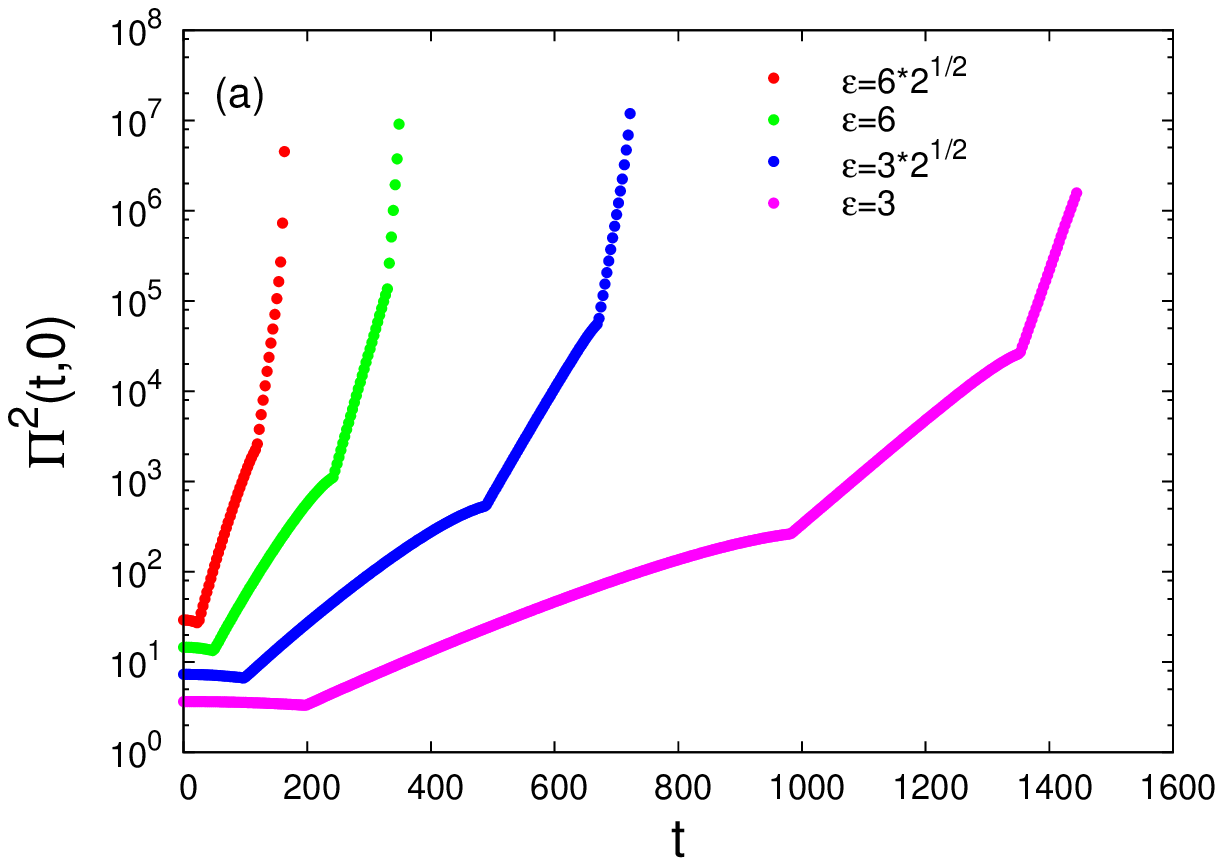}
  \includegraphics[width=0.48\textwidth]{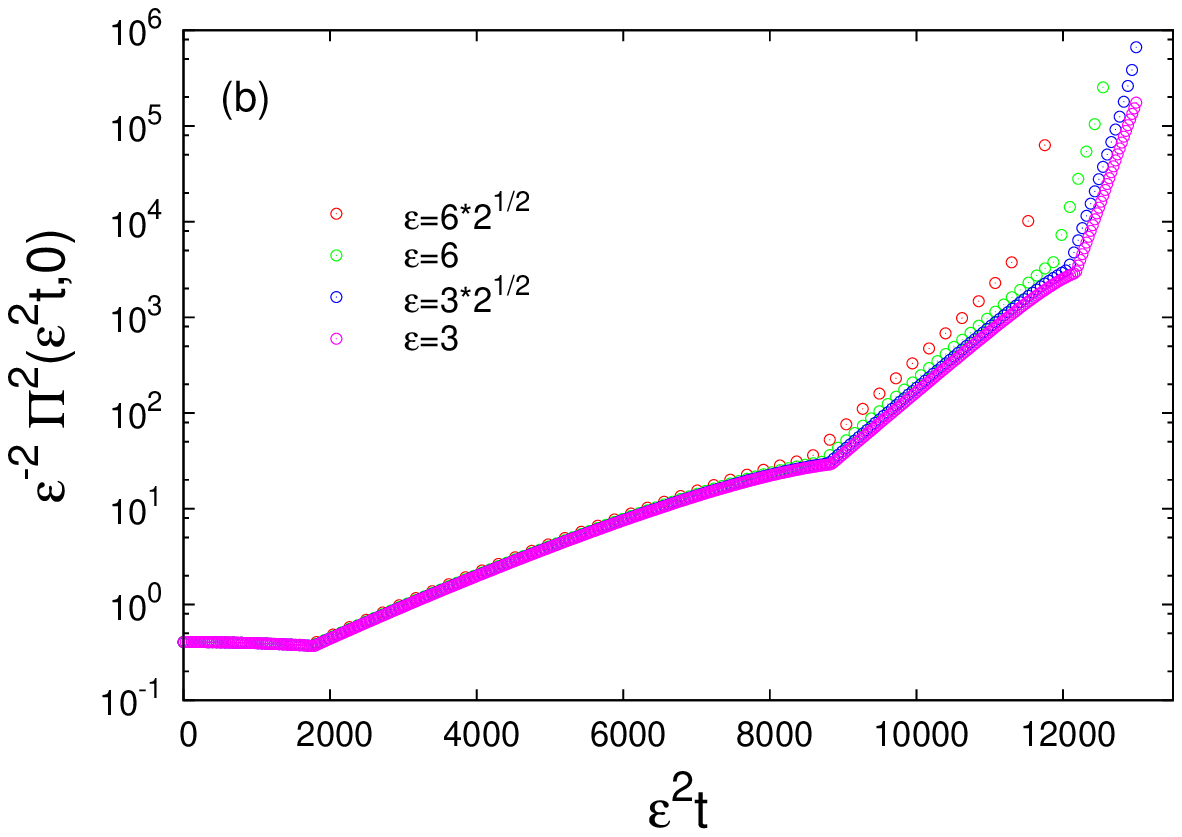}
   \caption{(a) $\Pi^2(t,0)$  for solutions with initial data \eqref{idata}
  for four moderately small amplitudes.
  For clarity of the plot only the upper envelopes of rapid oscillations are depicted.
  After making between about fifty (for $\ep=6 \sqrt{2}$) and five-hundred (for $\ep=3$) reflections, all solutions finally collapse.\\ (b) The curves from the plot (a) after rescaling $\ep^{-2} \Pi^2(\ep^2 t,0)$.}
  \label{fig2}
\end{figure}
\vskip 0.15cm
\noindent\emph{Weakly nonlinear perturbations.} We seek an approximate solution of the system (4-6) with initial data $(\phi,\dot\phi)_{|t=0}=(\ep f(x),\ep g(x))$ in the form
\begin{align}\label{pexp}
    \phi=\sum_{j=0}^{\infty} \phi_{2j+1} \ep^{2j+1}, \,\,
    A=1-\sum_{j=1}^{\infty} A_{2j} \ep^{2j},\,\, \delta=\sum_{j=1}^{\infty} \delta_{2j} \ep^{2j},
\end{align}
where $(\phi_1,\dot\phi_1)_{t=0}=(f(x),g(x))$ and all higher-order iterates have zero initial data.
Inserting \eqref{pexp} into the system (4-6) and collecting terms of the same order in $\ep$, we obtain a hierarchy of linear equations which can be solved order-by-order. At the first order we obtain
\begin{equation}\label{phi1}
    \ddot \phi_1 + L \phi_1=0,\quad
    L=-\frac{1}{\tan^2{\!x}}\, \partial_x \left(\tan^2{\!x} \,\partial_x\right)\,.
\end{equation}
This equation is a particular case of the  master equation describing the evolution of linearized perturbations of AdS space, analyzed in great detail by Ishibashi and Wald \cite{iw}.
The Sturm-Liouville operator $L$ is essentially self-adjoint on $L^2([0,\pi/2],\tan^2{\!x} \,dx)$.
 Below we denote the  inner product on this Hilbert space  by $(f,g):=\int_0^{\pi/2} f(x) g(x) \tan^2{\!x}\,dx$. The eigenvalues and eigenfunctions of $L$ are $\omega^2_j=(3+2j)^2$
 ($j=0,1,\dots$) and
\begin{equation}\label{modes}
e_j(x)=d_j \cos^3{\!x} \,{}_2 F_1(-j,3+j,\frac{3}{2};\sin^2{\!x})\,,
\end{equation}
where $d_j=\left(16(j+1)(j+2)/\pi\right)^{1/2}$ is the normalization factor ensuring that $(e_i,e_j)=\delta_{ij}$. The positivity of all the eigenvalues implies that AdS space is linearly stable. By elementary separation of variables, the solution of Eq.\eqref{phi1} is  given by the superposition of eigenmodes
\begin{equation}\label{phi1_exp}
    \phi_1(t,x)=\sum_{j=0}^{\infty} a_j \cos(\omega_j t+\beta_j)\, e_j(x)\,,
\end{equation}
where the amplitudes $a_j$ and  phases $\beta_j$ are determined by the initial data.

The back-reaction on the metric appears at the second order and can be readily integrated to yield
\begin{align}\label{second}
    A_2(t,x)&=\frac{\cos^3{\!x}}{\sin{x}} \int_0^x \left(\dot \phi_1(t,y)^2+\phi'_1(t,y)^2\right) \tan^2{\!y}\, dy,\\
    \delta_2(t,x)&=-\int_0^x \left(\dot \phi_1(t,y)^2+\phi'_1(t,y)^2\right) \sin{y}\cos{y}\, dy\,.
\end{align}

At the third order we get the inhomogeneous equation
\begin{equation}\label{phi3}
    \ddot \phi_3 + L \phi_3=S(\phi_1,A_2,\delta_2)\,,
\end{equation}
where $S:=-2(A_2+\delta_2)\ddot\phi_1-(\dot A_2+\dot \delta_2)\dot\phi_1 -(A'_2+\delta'_2)\phi'_1$.
Projecting Eq.\eqref{phi3} on the basis \eqref{modes} we obtain an infinite set of decoupled forced harmonic oscillations for the generalized Fourier coefficients $c_j(t):=(\phi_3,e_j)$
\begin{equation}\label{odes}
    \ddot c_j + \omega_j^2 c_j = S_j:=(S,e_j)\,.
\end{equation}
Let $I=\{j\in \mathbb{N}_0:a_j\neq 0\}$ be a set of indices of nonzero modes in the linearized solution \eqref{phi1_exp}. A lengthy but straightforward calculation yields that each triad $(j_1,j_2,j_3)\in I^3$ such that
$\omega_j=\omega_{j_1}+\omega_{j_2}-\omega_{j_3}$
 gives rise to a resonant term  in  $S_j$ (i.e., a term proportional to $\cos{\omega_j t}$ or $\sin{\omega_j t}$).
Some  of these resonances can be removed by a multiscale (or Poincar\'e-Lindstedt) technique, but for general initial data there is no way to remove all the resonances and consequently  we get secular terms which grow linearly in time and invalidate the perturbation expansion when $\ep^2 t=\mathcal{O}(1)$. For example, for the single-mode initial data, $(\phi,\dot\phi)_{t=0}=\ep(e_0(x),0)$, only $S_0$ contains a resonant term and this term can be eliminated by introducing a slow modulation of phase in \eqref{phi1_exp} to yield the approximation
$\phi(t,x)\simeq \ep \cos(3t+\frac{153}{4\pi} \ep^2 t)\,e_0(x)$ which is uniformly valid up to at least $t=\mathcal{O}(\ep^{-2})$. In contrast, for the two-mode initial data, $(\phi,\dot\phi)_{t=0}=\ep(e_0(x)+e_1(x),0)$,
the resonant terms in $S_0$ and $S_1$ can be eliminated by suitable phase modulations but the resonant term in $S_2$ persists and produces a secular term in $c_2(t)\sim t \sin{7t}$.

In accord with this analysis, we observe a striking difference in the evolution of these two kinds of solutions.
For (non-resonant) one-mode data the solution remains near the initial state during the entire simulation time, which indicates stability (although we cannot exclude a metastable behavior with a very long lifetime). In contrast,
for (resonant) two-mode data we observe the onset of exponential instability at time $t=\mathcal{O}(\ep^{-2})$, as in the case of gaussian-type data \eqref{idata}.
Notice that for $n$-mode initial data the number of  triads satisfying the diophantine resonance condition  grows rapidly with $n$ and hence, so does
the number of unremovable secular terms. We believe that these secular terms are progenitors of the higher-order resonant mode mixing which shifts the energy spectrum to higher frequencies.
 To quantify this effect, we introduce a spectral decomposition of the total energy as follows. We define the projections  $\Phi_j:=(\sqrt{A}\,\Phi,e'_j)$ and $\Pi_j:=(\sqrt{A}\,\Pi,e_j)$, and next, using the orthogonality relationships $(e'_j,e'_k)=\omega_j^2 \delta_{jk}$, we express the total conserved energy as the Parseval sum
$
    M=\frac{1}{2}\int_0^{\pi/2} \left(A\Phi^2+A\Pi^2\right) \tan^2{\!x}\,dx=\sum_{j=0}^{\infty} E_j(t)$,
where $E_j:=\Pi_j^2+\omega_j^{-2}\Phi_j^2$ can be interpreted as the $j$-mode energy \cite{mode}.
Numerical evidence corroborating the conjectured mechanism of instability is shown in Fig.~3
 which depicts the transfer of energy to higher frequencies  for a solution whose  energy is initially deposited in the first two modes, i.e. $E_j(0)\approx 0$ for $j\geq 2$.

  The  weakly nonlinear multiscale perturbation analysis of the system (4-6)
   will be elaborated in detail elsewhere \cite{br}, where we will also connect the nonlinear instability of generic quasiperiodic solutions of the linearized equations to the violation of non-resonance conditions of the KAM theorem for PDEs and Arnold diffusion.
\begin{figure}[h]
  \includegraphics[width=0.48\textwidth]{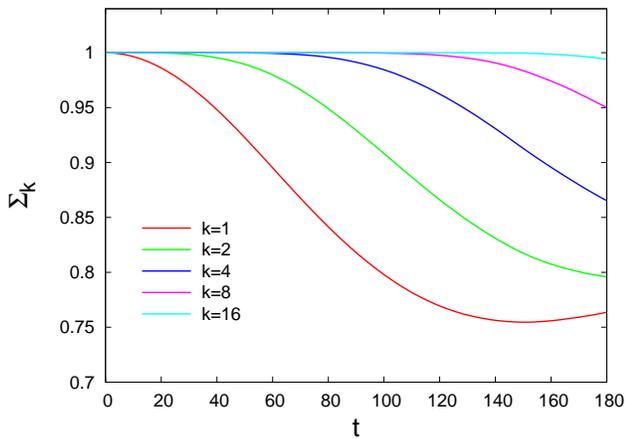}
  \caption{Evolution of the fraction of the total energy contained in the first $(k+1)$ modes
  $\Sigma_k:=\frac{1}{M}\sum_{j=0}^k E_j$ for the two-mode initial data
  $\phi(0,x)=\ep \left(\tfrac{1}{d_0} e_0(x)+\tfrac{1}{d_1} e_1(x)\right)$ with $\ep=0.088$.
  \vspace{-0.4cm}}
  \label{fig3}
\end{figure}
\vskip 0.15cm
\noindent\emph{Conclusions.}
There is growing (numerical and theoretical) evidence that the phenomenon of weak turbulence, i.e. the tendency of solutions to shift their energy from low to high frequencies,
 is common for (non-integrable) nonlinear wave equations on bounded domains, notably it
has  recently been proven for the nonlinear Schr\"odinger equation on torus \cite{c_team,cf}.
To our knowledge, this paper is the first result in this direction for Einstein's equations. An important difference between wave equations defined on a fixed background and Einstein's equations should be stressed. For the former, the weakly turbulent behavior is compatible with global-in-time smooth evolution. In contrast, for Einstein's equations the transfer of energy to high frequencies cannot proceed indefinitely because concentration of energy on  smaller and smaller scales inevitably leads to the formation of a black hole.

Admittedly, the results presented above raise more questions than give answers.  One of the key physical questions concerns  the role of negative cosmological constant $\Lambda$ in the observed phenomenon. Is it dynamical or kinematical? In other words, is an extra attractive force due to $\Lambda$  essential in triggering gravitational collapse
 for arbitrarily small perturbations, or
the only role of $\Lambda$ is to confine the evolution in an effectively bounded domain?

In this paper we studied perturbations of $AdS_4$ for the Einstein-massless-scalar system in $3+1$ dimensions, but we observed qualitatively the same behavior (in particular, instability of $AdS_5$) for the $4+1$ dimensional vacuum Einstein equations within the cohomogeneity-two biaxial Bianchi IX ansatz of \cite{bcs}. This result and its implications for the AdS/CFT conjecture will be discussed elsewhere.
\vskip 0.15cm \noindent \emph{Acknowledgments:} We thank Helmut Friedrich for helpful discussions and encouragement, and Piotr Chru\'sciel for critical reading of the manuscript. This work was supported in part by the NCN grant NN202 030740.


\begin{thebibliography}{10}
\bibitem{stability} In this Letter by stability we mean what is sometimes called nonlinear stability.
\bibitem{boundary} Although the proper distance from any point in the bulk to the boundary is infinite, from the point of view of wave dynamics, AdS is effectively bounded because  light rays reach infinity in finite time.
\bibitem{mode} Because of the nonlocal character of the total energy, this interpretation should be taken with a pinch of salt.
\end{thebibliography}
\end{document}